\documentclass{ws-p8-50x6-00}
\usepackage{url}

\newcommand{\email}[1]{{\ttfamily #1}}

\newcommand{\app}[3]{Astropart.\ Phys.\ {\bf #1} (#2) #3}

\newcommand{\prep}[3]{Phys.\ Rep.\ {\bf #1} (#2) #3}
\newcommand{\plb}[3]{Phys.\ Lett.\ {\bf B#1} (#2) #3}
\newcommand{\npb}[3]{Nucl.\ Phys.\ {\bf B#1} (#2) #3}

\newcommand{\apj}[3]{Astrophys.\ J.\ {\bf #1} (#2) #3}
\newcommand{\prl}[3]{Phys.\ Rev.\ Lett. {\bf #1} (#2) #3}
\newcommand{\prd}[3]{Phys.\ Rev.\ {\bf D#1} (#2) #3}

\newcommand{\href}[2]{#1}

\def\lsim{\mathrel{\rlap{\lower4pt\hbox{\hskip1pt$\sim$}}
    \raise1pt\hbox{$<$}}}         %less than or approx. symbol
\def\gsim{\mathrel{\rlap{\lower4pt\hbox{\hskip1pt$\sim$}}
    \raise1pt\hbox{$>$}}}         %greater than or approx. symbol

\begin{document}

\title{A new population of WIMPs in the solar system and indirect 
detection rates}

\author{Lars Bergstr{\"o}m}
\address{Department of Physics, Stockholm University, Box 6730,
SE-113~85~Stockholm, Sweden, \email{lbe@physto.se}}

\author{Thibault Damour}
\address{Institut des Hautes Etudes Scientifiques, 35 route de
Chartres, 91440 Bures sur Yvette, France,
\email{damour@ihes.fr}}

\author{\underline{Joakim Edsj{\"o}}\footnote{Rapporteur}}
\address{Department of Physics, Stockholm University, Box 6730,
SE-113~85~Stockholm, Sweden,
\email{edsjo@physto.se}}

\author{Lawrence M.\ Krauss}
\address{Departments of Physics and Astronomy,
Case Western Reserve
University, 10900 Euclid Ave, Cleveland OH 44106-7079,
\email{krauss@theory1.phys.cwru.edu}}

\author{Piero Ullio}
\address{Mail Code 130-33, California Institute of Technology,
Pasadena, CA 91125, USA,
\email{piero@tapir.caltech.edu}}


\maketitle

\abstracts{
A new Solar System population of Weakly Interacting Massive Particle
(WIMP) dark matter has been proposed to exist.  We investigate the
implications of this population on indirect signals in neutrino
telescopes (due to WIMP annihilations in the Earth) for the case when
the WIMP is the lightest neutralino of the MSSM, the minimal
supersymmetric extension of the standard model.  
The velocity distribution and capture rate of this new population is
evaluated and the flux of neutrino-induced muons from the center of
the Earth in neutrino telescopes is calculated.  We show that the
effects of the new population can be crucial for masses around 60--120
GeV, where enhancements of the predicted muon flux from the center of
the Earth by up to a factor of 100 compared to previously published
estimates occur.  As a result of the new WIMP population, neutrino
telescopes should be able to probe a much larger region of parameter
space in this mass range.  }

%%%%%%%%%%%%%%%%%%%%%%%%%%%%%%%%%%%%%%%%
\section{Introduction}

Weakly Interacting Massive Particles (WIMPs) can elastically scatter
inside the Sun and Earth, leading to their subsequent capture and
annihilation in the cores of these bodies, producing an indirect
neutrino signature that might be accessible to neutrino
telescopes.\cite{genrefs} Recently it has been demonstrated that the
scattering process in the Sun can populate orbits which subsequently
result in a bound Solar System population of WIMPs \cite{dk1,dk2} and
which can be comparable in spectral density, in the region of the
Earth, to the Galactic halo WIMP population.  This new population
consists of WIMPs that have scattered in the outer layers of the Sun
and due to perturbations by the other planets (mainly Jupiter and
Venus) evolve into bound orbits which do not cross the Sun but do
cross the Earth's orbit.  This population of WIMPs should have a
completely different velocity distribution than halo WIMPs and will
thus have quite different capture probabilities in the Earth.  The
predicted WIMP abundance, and spectrum, relevant for direct detection
have been calculated \cite{dk1,dk2} and here we focus on capture in
the Earth, and the predicted indirect neutrino signature.

Following \cite{dk1,dk2}, we consider here a special WIMP candidate,
the neutralino, which arises naturally in supersymmetric extensions of
the standard model.  Although our numerical results apply only to
supersymmetric WIMPs, we expect our qualitative conclusions to be
generally valid for any WIMP model of dark matter. 

For more details on this work, see \cite{dkpop}.

%%%%%%%%%%%%%%%%%%%%%%%%%%%%%%%%%%%%%%%%
\section{The MSSM}

We evaluate the capture of neutralinos and the resulting
neutrino-induced muon flux within the Minimal Supersymmetric extension
of the Standard Model (MSSM) (for a review of neutralino dark matter,
see \cite{jkg}).  For the numerical MSSM calculations we use
{\sffamily DarkSUSY} \cite{darksusy} with which we have generated about 
140\,000 different MSSM models and for each one we require that they 
do not violate current accelerator constraints and that they make up a 
major part of the dark matter, i.e.\ $0.025<\Omega_{\chi}h^{2}<1$, 
with $\Omega_{\chi}$ being the neutralino relic density and $h$ being 
the Hubble constant in units of 100 km s$^{-1}$ Mpc$^{-1}$.

%%%%%%%%%%%%%%%%%%%%%%%%%%%%%%%%%%%%%%%%
\section{Density and velocity of the new WIMP population}

\begin{figure}
\begin{minipage}[t]{0.48\textwidth}
\centerline{\epsfig{file=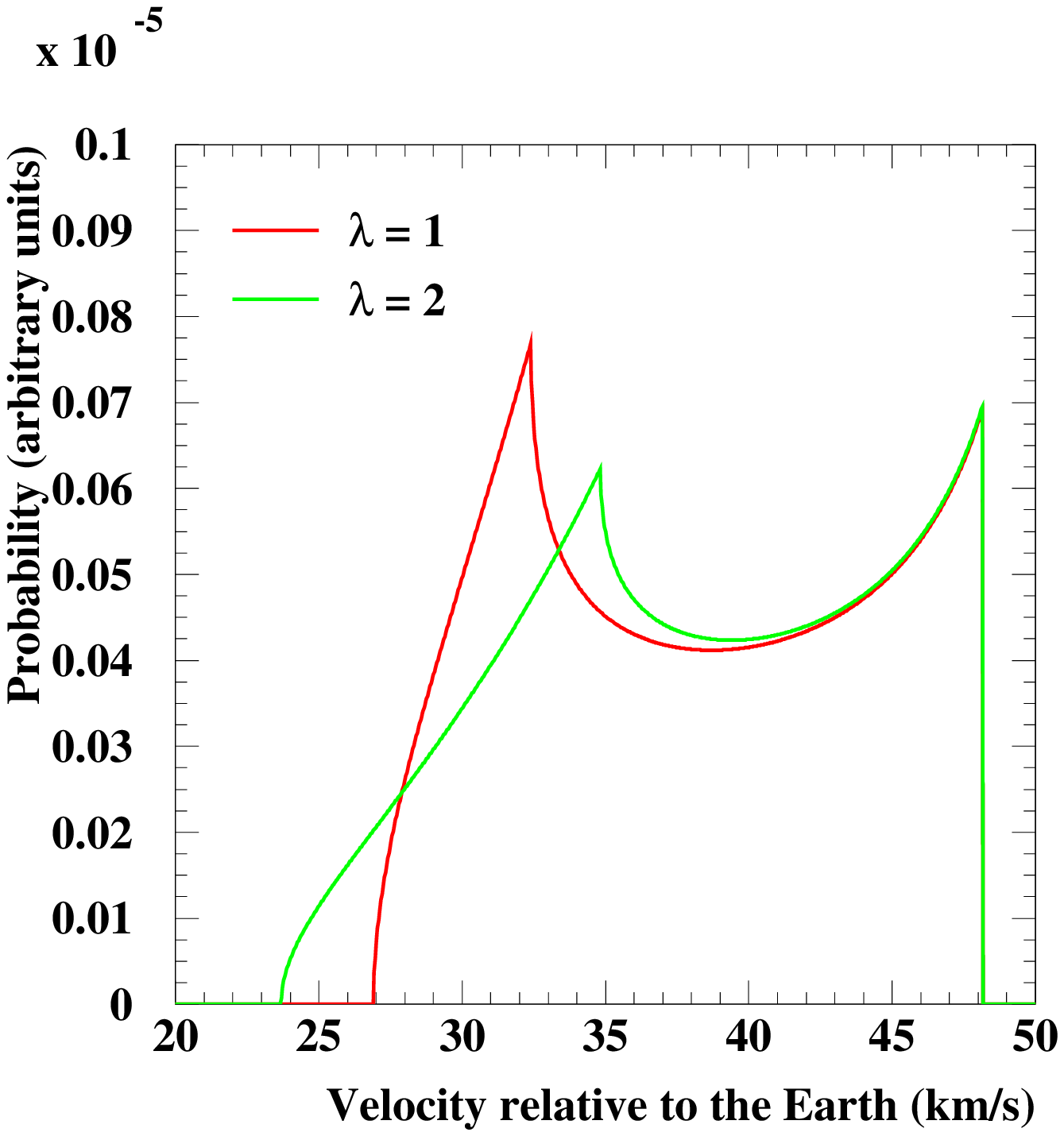,width=\textwidth}}
\caption{The velocity distribution with respect to the Earth for the new 
population of neutralinos.}
\label{fig:vdist}
\end{minipage}
\hfill
\begin{minipage}[t]{0.48\textwidth}
\centerline{\epsfig{file=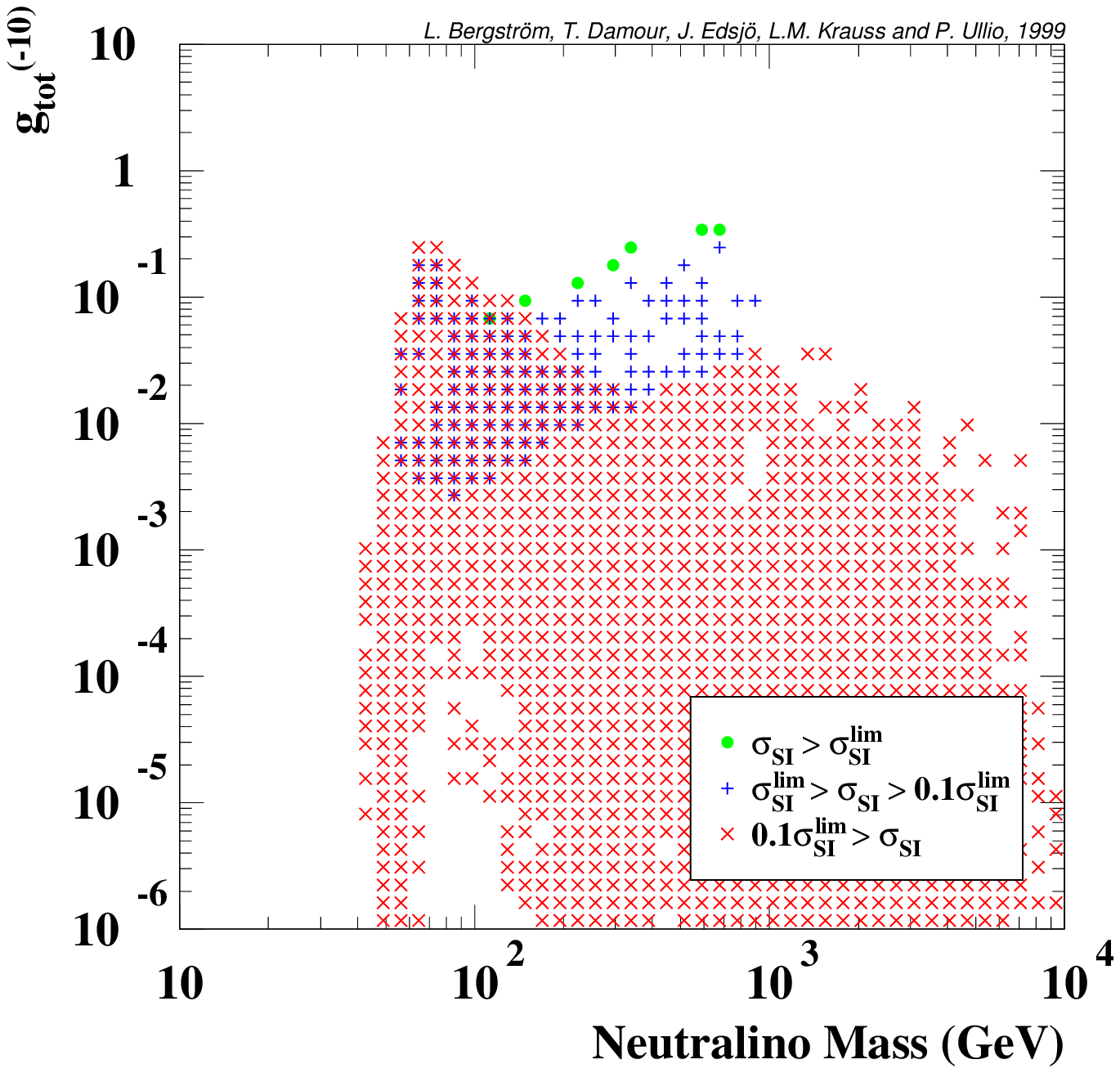,width=\textwidth}}
\caption{$g_{\rm tot}^{(-10)}$, which is related to the overdensity is 
shown for the new population of neutralinos.}
\label{fig:gtot}
\end{minipage}
\end{figure}

WIMPs that scatter in the outskirts of the Sun can end up in highly
elliptical orbits gravitationally bound to the Sun.  Jupiter and/or
Venus can perturb these orbits so that they no longer cross the Sun,
but if the semi-major axis is in the range $0.5 a_{\rm earth} < a <
2.6 a_{\rm earth}$ they do cross the Earth's orbit and can be captured
by the Earth while they still don't reach out to Jupiter in which case 
they would be gravitationally scattered away.\cite{dk1,dk2} 

The orbits will be nearly radial and the velocity with respect to the
Earth is between 27 km/s and 48 km/s.  Based on a detailed
calculation\cite{dkpop}, we show the velocity distribution of these
neutralinos in Fig.~\ref{fig:vdist}.  $\lambda$ is a parameterization
of the non-conservation of the $z$-component of the angular momentum
(if the orbit is in the $xy$-plane.  $\lambda=1$ corresponds to full
conservation of $J_{z}$ and $\lambda=2$ corresponds a non-conservation
by a factor of two.  We have checked that our results not depend much on
$\lambda$ and we will in the following focus on $\lambda=1$.

In \cite{Gould91}, the phase space distribution of WIMPs at the Earth
was investigated including the effects of the Earth being deep in the
Sun's potential well.  There it was found that even though the
velocity distribution of WIMPs at the Earth is different than it would
be in free space, both Jupiter, the Earth and Venus will disturb the
orbits of these unbound WIMPs into bound orbits that have the same
phase space distribution as would be the case in free space.  This
transfer in phase space is only fast enough (compared to the age of
the solar system) for low velocities, $\lsim$ 27 km/s. For our new 
population of WIMPs, which has higher velocities, this means that they 
are not diffused out of the solar system by gravitational interactions 
with the other planets. The scatterings in the outskirts of the Sun will 
thus have built up this new population since the solar system was 
formed.

We can write the enhancement of the local neutralino density 
from this new population as
\begin{equation}
  \delta_E \equiv \frac{\hbox{WIMP density from new population}}
  {\hbox{WIMP density from old population}} 
  =   \frac{0.212}{(v_o / 220 \,
  {\rm km\,s}^{-1})} \, g_{\rm tot}^{(-10)}.
  \label{eq:deltae}
\end{equation}
Here, $g_{\rm tot}^{(-10)}$ is a quantity that depends on the details 
of the scattering of WIMPs in the outskirts of the Sun and on the 
perturbation of the orbits from mainly Jupiter and Venus. $v_{o}$ is 
the local rotation velocity around the galactic center. In 
Fig.~\ref{fig:gtot} we show the values for $g_{\rm tot}^{(-10)}$ 
evaluated for our MSSM models. Using Eq.~(\ref{eq:deltae}) we see 
that the overdensity is up to $\delta_{E} \simeq 0.08$. For the local 
halo density of neutralinos we have used 0.3 Gev/cm$^{3}$.

%%%%%%%%%%%%%%%%%%%%%%%%%%%%%%%%%%%%%%%%
\section{Capture rate of the new population}

Even though the density of the new population is at maximum 10\% of
the halo density, the velocity distribution is quite different.  Since
the capture efficiency in the Earth is highly velocity dependent we
would expect the capture rate to be quite different from the normal
scenario.  For capture in the Earth, scattering off iron nuclei is the
most efficient process and given the lower limit of the velocity
distribution in Fig.~\ref{fig:vdist}, we can calculate that capture of
the new population is only possible when the neutralino mass,
$m_{\chi} \le 2.90 m_{Fe} \simeq 150$ GeV\@.

For our set of MSSM models, we have calculated the capture rate 
from this new population and derived the neutralino annihilation rate 
by solving the evolution equation for neutralinos in the Earth 
including both the old and the new population of 
neutralinos.\cite{dkpop} We then define the enhancement of the 
annihilation rate (and hence on the resulting neutrino-induced muon 
fluxes) from this new population as
\begin{equation}
    {\cal E} \equiv \frac{\Gamma_{a}^{\rm tot} - \Gamma_{a}^{\rm old}}
    {\Gamma_{a}^{\rm old}} =
    \frac{\Phi_{\mu}^{\rm old+new}-\Phi_{\mu}^{\rm old}}
    {\Phi_{\mu}^{\rm old}}.
    \label{eq:enh}
\end{equation}
In Fig.~\ref{fig:enhancement} we show the enhancement factor for our MSSM
models, and as we see, the enhancement can be up to two orders of
magnitude when the mass of the neutralino is in the region 60--120 GeV
where scattering off iron is efficient.

%%%%%%%%%%%%%%%%%%%%%%%%%%%%%%%%%%%%%%%%
\section{Neutrino-induced muon fluxes}

\begin{figure}
\begin{minipage}[t]{0.48\textwidth}
\centerline{\epsfig{file=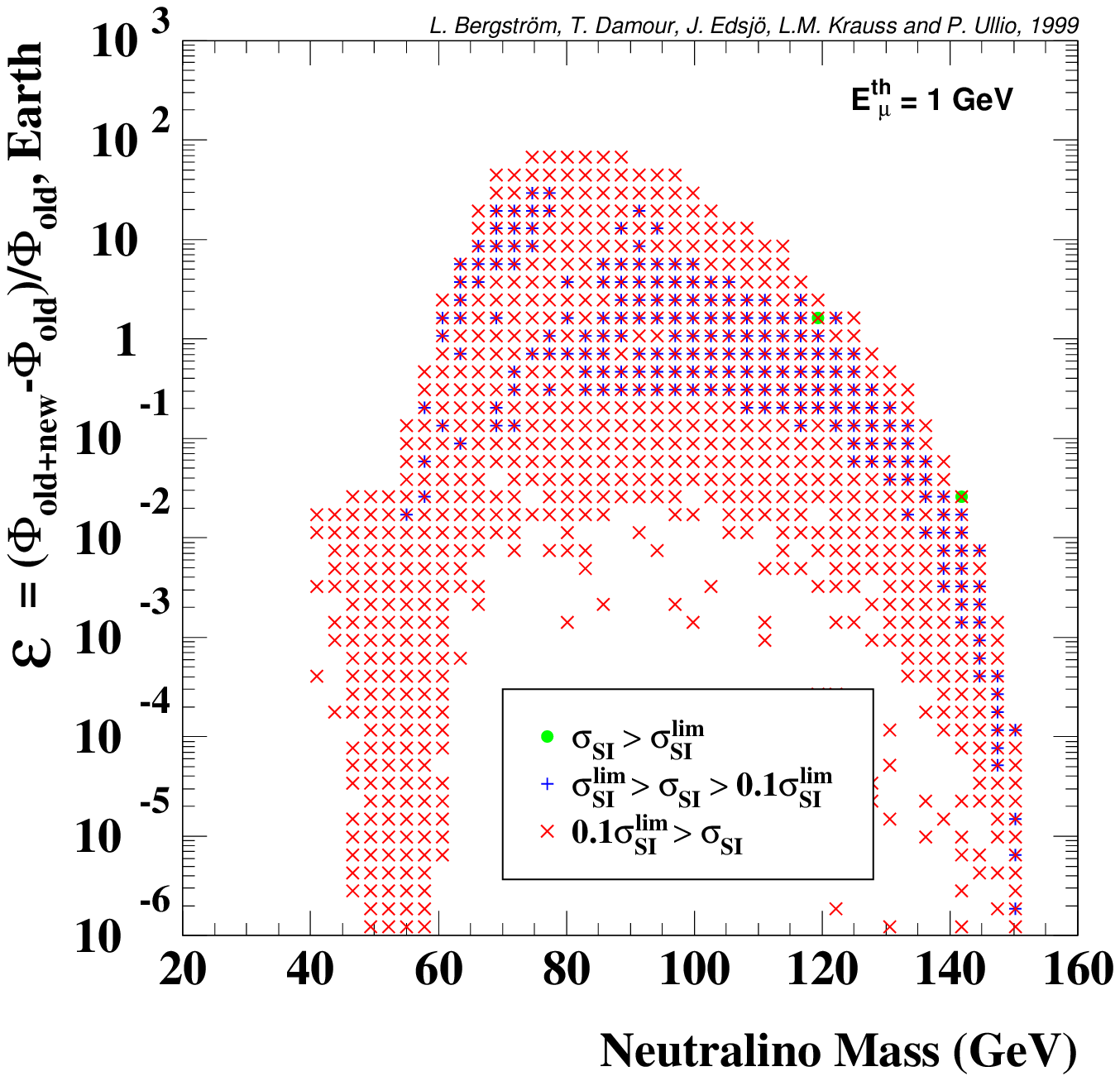,width=\textwidth}}
\caption{The enhancement factor from the new population of neutralinos.}
\label{fig:enhancement}
\end{minipage}
\hfill
\begin{minipage}[t]{0.48\textwidth}
\centerline{\epsfig{file=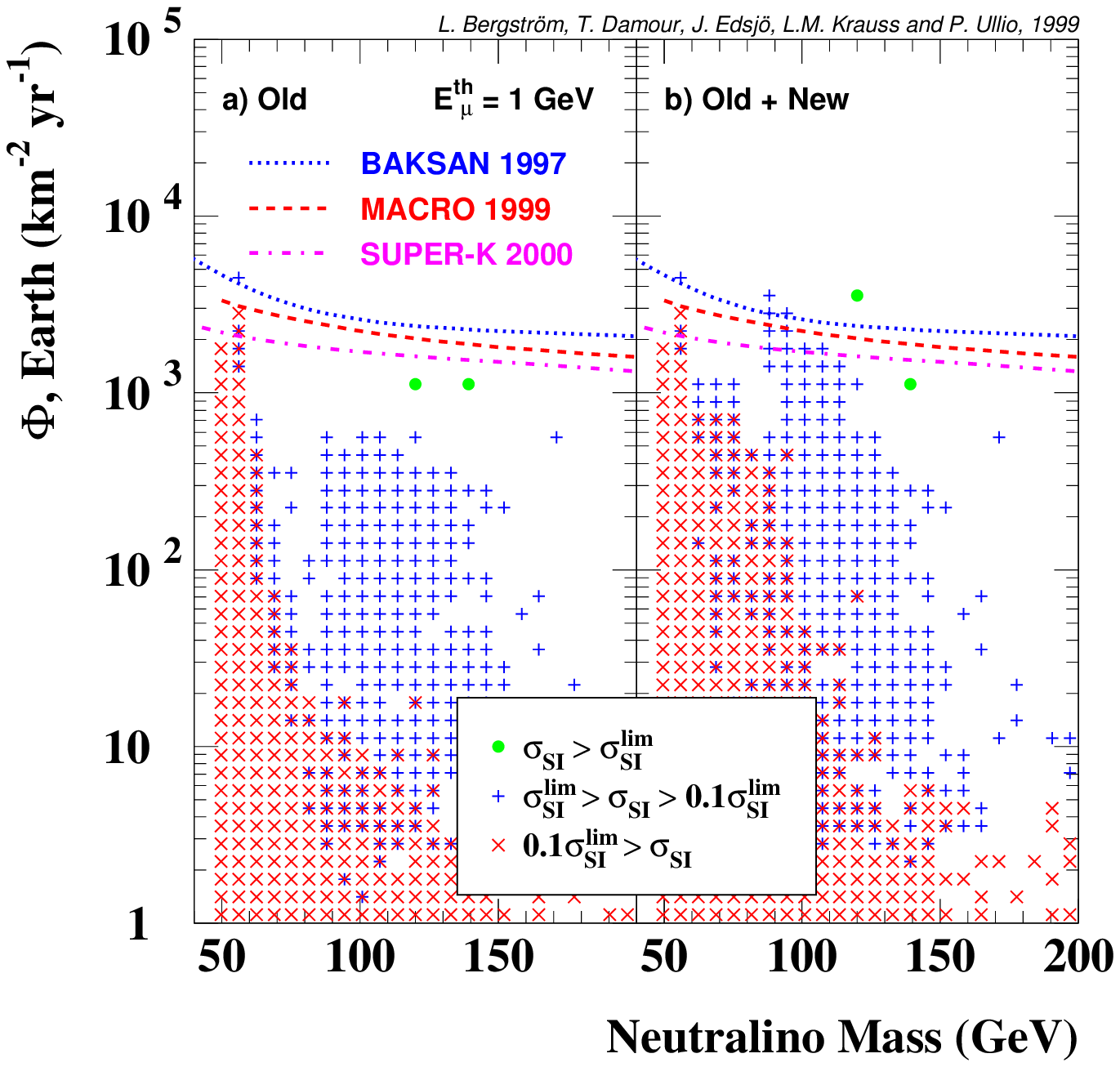,width=\textwidth}}
\caption{The 
neutrino-induced muon fluxes are shown for a) the old population and 
b) the old and the new population of neutralinos.}
\label{fig:muflux}
\end{minipage}
\end{figure}

We are now ready to compare the absolute fluxes of neutrino-induced
muons from neutralino annihilation in the center of the Earth when we
include or not include the new population of neutralinos.  We
calculate the neutrino-induced muon flux as described in \cite{km2}
and in Fig.~\ref{fig:muflux} we show the fluxes above 1 GeV without
and with the new population included.  We see that for the models with
the highest fluxes, we get an enhancement of about an order of
magnitude when the neutralino mass is 60--120 GeV\@.  We also show the
current limits from Baksan\cite{baksan}, Macro\cite{macro} and
Super-Kamiokande\cite{superk} and we see that these experiments are
much more sensitive to MSSM models in this mass range when this new
population of neutralinos is included.

%%%%%%%%%%%%%%%%%%%%%%%%%%%%%%%%%%%%%%%%
\section{Conclusions}

We have investigated a new population of WIMPs in the solar system
which arise when WIMPs that have scattered in the outskirts of the Sun
are trapped in orbits gravitationally bound to the Sun, but due to
perturbations by mainly Jupiter and Venus no longer intersect the Sun. 
This new population of WIMPs have nearly radial orbits and can cross
the Earth's orbit and be captured by the Earth.  There they can
annihilate pair-wise producing muon neutrinos that can be searched for
with neutrino telescopes.  We have focused on when the WIMP is a
neutralino in the MSSM and have shown that the fluxes can be enhanced
by up to a factor of 10 when the neutralino mass is in the range
60--120 GeV\@. Hence, this significantly enhances the sensitivity for 
these experiments to neutralino dark matter.

%%%%%%%%%%%%%%%%%%%%%%%%%%%%%%%%%%%%%%%%
\section*{Acknowledgments}
L.B.\ was supported by the Swedish Natural Science Research Council
(NFR).  T.D.\ was partially supported by the NASA grant
NAS8-39225 to Gravity Probe B (Stanford University).
The research of L.M.K. was supported in part by the U.S.
Department of Energy.

\end{document}